\documentclass{article}

\usepackage{graphicx}
\usepackage{amsmath}
\usepackage{amssymb}
\usepackage{booktabs}
\usepackage{float}

\def\nsf{\textit{Next Step Fusion s.a.r.l.}}
\title{Reinforcement Learning Controllers for Fusion Scenario Development}
\author{Evgeny Adishchev, Maxim Nurgaliev, Dmitry Sorokin,\\Alexander Granovsky,
Maksim Stokolesov, Igor Prokofyev,\\Eduard Khairutdinov, Georgy Subbotin\\\nsf}
\date{October 6, 2026}

\emergencystretch=\maxdimen
\begin{document}

\maketitle


\begin{abstract}
Scenario development, the design of the waveforms that take a tokamak discharge from
breakdown to flat-top, is still an iterative, labour-intensive process. Every candidate
scenario has to be simulated with a controller in the loop, and the classical shape
and current controllers need low-level actuator references. So the
process is iterative and includes a number of trial-and-error steps.
We propose to replace the per-scenario controller tuning by a
single goal-conditioned reinforcement-learning controller trained once, in advance,
against a free-boundary equilibrium and transport simulator. A scenario is then
specified only by the target waveforms of the physical quantities of interest (plasma
current, line-averaged electron density, LCFS shape, x-point positions), and the controller
produces the coil-current and fueling commands that realise them. We demonstrate the
approach on an EAST plasma-current ramp-up: the controller
takes over a small limited plasma, ramps the
current and the density, grows the cross-section, forms two x-points and holds a
double-null diverted configuration against sampled shape targets for the rest of a
1.2~s episode. The neural network controller is trained using TQC algorithm. On 1024 test
scenarios sampled from a low-discrepancy scrambled Sobol sequence in
11-dimensional target space, the controller forms the divertor
in 86\% of the cases, holds it for more than 0.5~s in 67\% and reaches the end of the
episode in a diverted state in 58\%, with flat-top tracking errors of a few kA in
plasma current, $\sim 0.15\times10^{19}$~m$^{-3}$ in density, $\sim 2$~cm in position
and $\sim 8$~cm in x-point location; the failures concentrate at the high-current density
($J$) region which is believed to be physically or from engineering perspective
not feasible within ohmic mode. 
\end{abstract}

\section{Motivation}
\label{sec:motivation}

This paper presents a tokamak operating scenario as the time history of the MHD states,
coils' currents, other actuators including fueling and heating from breakdown to the flat-top and
later to the ramp down. Its
development is one of the most time-consuming activities in the preparation of an experimental campaign,
as it is largely an unstructured trial-and-error
process~\cite{na2025,humphreys2015,detommasi2019}. The scenario development is done with integrated simulators,
those inputs are actuator values and outputs---plasma states evolved over a discharge.
In operations the actuators depend on the plasma state and determined by a controller---typically
PID using specific set points. During scenario design and optimization the set points
are not known yet, the scenario designer therefore iterates:
propose waveforms, re-tune the controller, simulate.
Exploratory optimisation is particularly limited---small modifications
frequently lose the equilibrium or hit a coil limit, and each failed attempt costs a
simulation and a manual diagnosis.

Our proposal is to run these simulations in advance. Instead of tuning a controller
for every candidate scenario we train, once, a goal-conditioned controller that receives
the target waveforms of the physical quantities of interest together with the plasma
state (MHD and kinetic) and produces the actuator commands. Once such a controller exists, a scenario is
specified entirely in physical terms---``ramp to 300~kA in 0.6~s, reach
$\bar n_e = 2\times10^{19}$~m$^{-3}$, form a double-null shape with elongation 1.9 and
these x-point radii''---and each iteration of the design loop is a single closed-loop
simulation that is, by construction, consistent with the plant model. The capable
simulator intrinsically takes engineering boundary conditions into accout and
produces a scenario that fits for purpose and also feasible to the device. Variations can be
explored interactively, and the controller's own test statistics tell the designer
which region of the target space is reachable.

The dimensionality of the problem rules out exhaustive search: the state of a
free-boundary equilibrium coupled to a transport model has hundreds of degrees of
freedom, the actuator space has about a dozen continuous channels, and a scenario is a path
through this space over a thousand control steps. Approximate, learned solutions are
needed, and deep reinforcement learning (RL) using neural networks as
universal approximators is the natural tool for the complex system
with nonlinear evolution and control dynamics: it optimises a
closed-loop policy directly against a simulator, handles continuous multi-input
multi-output control, and has already been shown to control tokamak plasmas in
simulation and on real devices. Magnetic control by RL has been demonstrated on
TCV~\cite{degrave2022}, DIII-D~\cite{subbotin2026,sorokin2026},
EAST~\cite{zhang2026}, HL-3~\cite{hl3_2024} and
EXL-50U~\cite{ding2026,guo2026}; kinetic and 0-D scenario quantities have been
controlled on KSTAR~\cite{seo2021,seo2022} and JT-60SA~\cite{wakatsuki2023}, and RL has
been used for proactive instability avoidance on DIII-D~\cite{seo2024} and for robust ramp-down
trajectory design on TCV~\cite{wang2025}.
Most of these works controls a plasma that
already exists in its target configuration (static shape hold, disturbance rejection,
profile regulation) or produces feed-forward trajectories. The present work addresses
the scenario itself: the controller has to carry the plasma through the qualitatively
different limited and diverted phases of a ramp-up, with the limiter-to-divertor
transition---a discrete event that classical control systems handle by switching
control laws---learned as part of the policy. We focus on ramp up to flat-top
transition excluding breakdown itself and ramp down.

For this paper we applied the method to EAST device and produced a controller
for offline use. The controller develops scenarios in a wide range of target shapes
and kinetic characteristics. This range covers a substantional subspace of
double null configurations achievable without external heating.

\section{Scenario Development as a Control Task}
\label{sec:control-task}

To an operator, a scenario is a small set of time histories of physical quantities: the
plasma current ($I_P$), the line-averaged electron density ($n_e$), and a parametrised
description of the plasma
boundary---the inner-wall gap, the minor radius, the elongation, the triangularity, the squareness,
the positions of two x-points.  Between the scenario and the machine sits
a controller that has to turn the desired history of the plasma states into a history of
power system and fuelling commands.

Such a controller must be \emph{goal-conditioned}: the target parameters enter as inputs, so that the
same controller serves every scenario in the family. The
targets move throughout the discharge, and the controller has to track ramps rather
than regulate about a fixed point. And it must handle the \emph{phase structure} of a
ramp-up on its own: a scenario begins with a small plasma leaning on the inner wall and
ends with a diverted plasma whose x-points are themselves tracked quantities; the
targets that exist in the second phase do not exist in the first, the failure modes
differ, and the transition between the two is a discrete event that has to happen
at the right moment (chosen by the controller) and from the right state.

We cast the problem as an episodic RL task. An episode starts a few
tens of milliseconds after breakdown from a realistic early-discharge equilibrium and
lasts a fixed physical time, we used 1.2~s. Targets values are determined as piece wise linear
quantities starting from the initial state (at $T=0$ control error = 0), going
at a constant pace to the final values and at a certain moment change to constants.
The change to constant phase may vary between targets and when all of them
are at the constant phase, this is a discharge flat top.
The controller observes the
plasma, the coil currents and the current targets at every millisecond and commands
changes of the coil currents and of the fueling rate. It is rewarded at every step for
how close the achieved quantities are to their targets, and the episode ends early if
the equilibrium is lost, if an operational limit is violated, or if a tracked quantity
leaves a tolerance band. 

Three properties of ramp-up physics make this harder than the static control tasks
addressed by the earlier RL work and shape the design choices of the next section.
First, the current, the density and the shape are coupled through transport. Second, the
actuators act on different time-scales: a coil current changes within milliseconds,
the density responds to the gas valve over tens to hundreds of miliseconds, so the effect of a
fueling decision is only visible far beyond the horizon over which a learning algorithm
naturally assigns credit. Third, and most importantly, the decisive event of the
scenario---a limiter-to-divertor transition that survives---only occurs after a few
hundred steps of control of the limited plasma, and a controller that never
attempts it is rewarded comfortably for holding a stable limited plasma. Discovering
the transition is as much an exploration and optimization problem as a control problem.

\section{RL Algorithm Set-up (MDP)}
\label{sec:mdp}

\subsection{Environment}
\label{sec:env}

We used NSFsim, a free-boundary Grad--Shafranov equilibrium solver coupled to
a one-dimensional transport model~\cite{clark2025} which is able to simulate
any axisymmetric fusion device. The device description---coils, vessel, limiter,
magnetic probes and flux loops---is configured separately, in the control
task we assume that the simulator is verified.
The results below use the EAST 2021 configuration: major radius
1.85~m, toroidal field 1.8~T on axis, twelve poloidal-field coils (PF1--PF8 and
PF11--PF14, of which PF1--PF6 form the central solenoid), one in-vessel coil (IC) and a
90-element model of the conducting vessel. The equilibrium and the transport equations
are advanced with a time step of 1~ms.

Every episode starts from the same reconstructed early-discharge equilibrium: a
circular plasma limited on the inner wall with minor radius 0.20~m, $I_P \approx 41$~kA
and the magnetic axis at the midplane. The vertical stability is controlled with the conventional vertical
stability controller which calculates the compensation $B_R$ field and applies it
through the IC coil. Everything else is under the policy's control from the first step. The control
step is the same 1~ms and the episode is truncated at 1.2~s, i.e.\ 1200 control steps,
unless it terminates earlier.

The controller commands are \emph{changes} of the currents in PF1--PF8 and PF11--PF14.
Three up--down pairs of the central-solenoid coils are driven symmetrically
(PF2 with PF1, PF4 with PF3, PF6 with PF5), which leaves nine independent
coil channels. The change per step is bounded by
the supply voltage limit and the coil self-inductance, and
the currents themselves by the rated 12.5~kA, so the power-supply capability enters
through the action space. The tenth channel is the neutral
gas source, parametrised as a rate-limited change of a fueling level that is converted
into an edge-localised deuterium particle source. In the same manner as for the coils
the policy output is a change of the fuelling rate from the previous step.

The controller observes the quantities a shape and current controller would be given by
a real-time equilibrium reconstruction, plus the targets: the plasma current and the
coil currents with their time derivatives. The target shape is parametrized by
magnetic center R position, minor radius, inner-wall gap, elongation, coordinates of
x-points and four values of squareness. The model is fed
by the eight pivot points instead of mixed length and unitless quantities. These are
right, left, top, bottom points of the separatrix plus four points which are determined
by the squareness for each quadrant. Additional observations are
the flux imbalance between the two nulls; a few scalar equilibrium and
profile quantities. The full list is given in~\ref{app:obs}.

Table~\ref{tab:targets} lists the ranges from which the scenario targets are drawn.

\begin{table}
\caption{\label{tab:targets}Scenario target ranges, sampled uniformly and independently
at every training episode. Each target is a ramp from the value measured at hand-over to
the sampled final value, reached at the sampled time.}
\centering
\begin{tabular}{@{}llll}
\toprule
Quantity & Symbol & Final value & Ramp end [s]\\
\midrule
Plasma current & $I_p$ & 200--350 kA & 0.4--1.0\\
Line-averaged density & $\bar n_e$ & 1.5--3.0 $\times10^{19}$ m$^{-3}$ & 0.4--1.0\\
Inner-wall boundary radius & $R_{\rm in}$ & 1.40--1.53 m & 0.4--1.0\\
Minor radius & $a$ & 0.24--0.36 m & 0.4--1.0\\
Elongation & $\varkappa$ & 1.7--2.0 & 0.4--1.0\\
Squareness (4 quadrants) & $\zeta$ & 0.2--0.4 & 0.4--1.0\\
Lower / upper x-point radius & $R_{x1}$, $R_{x2}$ & 1.53--1.63 m & 0.5--1.0\\
Vertical position of the centre & $Z_{\rm mag}$ & 0 (fixed) & 0.03\\
\bottomrule
\end{tabular}
\end{table}

\subsection{Reward Engineering}
\label{sec:reward}

The reward contains several components lying in $[0; 1]$.
Each reward component starts with some scalar error $d_k(g, s) \ge 0$---a distance between the
target $g$ and the state $s$---with a transform onto $[0,1]$. The transform is the
Lorentzian
\begin{equation}
r_k = \frac{1}{1 + (d_k/\gamma_k)^2},
\end{equation}
so that if $\gamma_k$ is the error the component is worth one half and
$\frac{1}{3}\gamma_k$ delivers the reward of $0.9$.

The components are combined into a single value with a Boltzmann-weighted smooth minimum that lies
between the smallest and the largest component
\begin{equation}
R = \sum_k w_k r_k, \qquad w_k = \frac{\exp(\beta r_k)}{\sum_j \exp(\beta r_j)}, \qquad
\beta = -5,
\label{eq:softmin}
\end{equation}

With $\beta<0$ the combined reward is dominated by the worst component. This is a deliberate
choice: a plain weighted sum lets the policy trade one objective for another, whereas the
smooth minimum only pays for improving whichever objective is currently worst, so a converged policy
satisfies all of them to a comparable degree.

The components fall into three groups. Tracking components measure the distance of the
plasma current, the line-averaged density, the geometric centre, the eight pivot
points (reduced to a scalar by the same formula \ref{eq:softmin}, with positive $\beta=5$
it turns into the smooth maximum instead if minimum), so that the worst point drives the
component without the discontinuity of a hard maximum), the two x-points, the flux
imbalance between the two nulls ($|\psi(x_1) - \psi(x_2)|$), and the normalised
poloidal flux at the eight
pivot points from their targets. Operational-limit components price the spread of
the six central-solenoid currents, the current in the in-vessel IC
coil as a proxy to vertical stability, and the margin of every coil to its
max current. A limiter component rewards a
contact point on the midplane, i.e.\ a plasma that leans on the inner wall symmetrically
before it is diverted. 

The episodes are phased. Phase 0 is the limited plasma, and the phase 1 is the
diverted plasma. The phase progression happened when the controller collects
10~observations of the diverted plasma (regardless in a row or in a flickering manner)
with $I_P > 100$~kA and also if the termination criteria for the phase 1 is
not met (see below). The phase is (0 or 1) is added to the reward on top of
smooth minimum of the components. This design sends a strong signal to the algorithm
to transit to the diverted mode, allowing a short deviation of other targets
(namely shape and position) for a higher reward value in several dozens of control
cycles.  Some reward components are phase specific. If a component is not used in
a phase, it does not enter the smooth minimum averaging.

\subsection{Episode termination}
An episode terminates when some criteria is met. The first series of criteria is
plasma survival: the MHD equilibrium existence, the $I_P$ being above 20~kA
 regardless of the target, the minor radius to be above 0.2~m and the
edge safety factor above $2.0$.

Another set of criteria is tied to the target. An episode is terminated if
pivot points deviate more than 0.2~m to their target, $I_P$---more than 50~kA etc.

The third series of criteria is phase 1 (the diverted plasma) specific. The flux
imbalance between x-points must be below 0.1 and x-point distance to target must be under 0.2~m
during phase 1. Touching a wall causes termination as well.

And lastly there is an engineering criterion, the CS spread (maximum current minus
minimum current) must not grow more than 9.5~kA off the starting value. This
focuses the model to use CS for the plasma current control only and not
going to different direction effectively using PF5-6 for shaping.

Maximum length of an episode is 1.2~s, i.e. 1200 steps. Reaching 1200 steps is
marked as truncation rather then termination and not penalized.

\subsection{RL Algorithm}
\label{sec:algo}

We use an off-policy maximum-entropy actor--critic. The actor is a Gaussian policy with
a $\tanh$ squashing, trained with the soft actor--critic objective and automatic
temperature tuning~\cite{haarnoja2018}; the critic is the truncated
quantile ensemble of TQC~\cite{kuznetsov2020}: three networks each predicting 25
quantiles of the return distribution, whose pooled atoms are sorted and the largest few
dropped before the bootstrap target is formed. Both networks are two-layer perceptrons
of width 256 with SiLU activations on top of a running input normaliser (Welford's
algorithm). The critic additionally carries an auxiliary head that predicts the shape
scalars (triangularity, elongation and squareness) which are not present in
the observations directly; its mean-squared loss, added to the policy loss with a small
weight, provides a training signal while successful experience is not yet collected.
The discount factor of 0.99 gives
an effective horizon of 100~ms, long enough to cover the vessel eddy-current response
and the coil slew limits. Learning rates, batch and replay sizes and the
remaining hyperparameters are listed in~\ref{app:hyper}.

\section{Results}
\label{sec:results}

The controller reported below is trained with the configuration of
Section~\ref{sec:mdp} on the target ranges of table~\ref{tab:targets}. All the
results of this section are obtained with a single model, the checkpoint after
about 4.0~million updates, run as a deterministic policy.

\subsection{The developed controller}
\label{sec:controller}

\begin{figure}
\centering
\includegraphics[width=\textwidth]{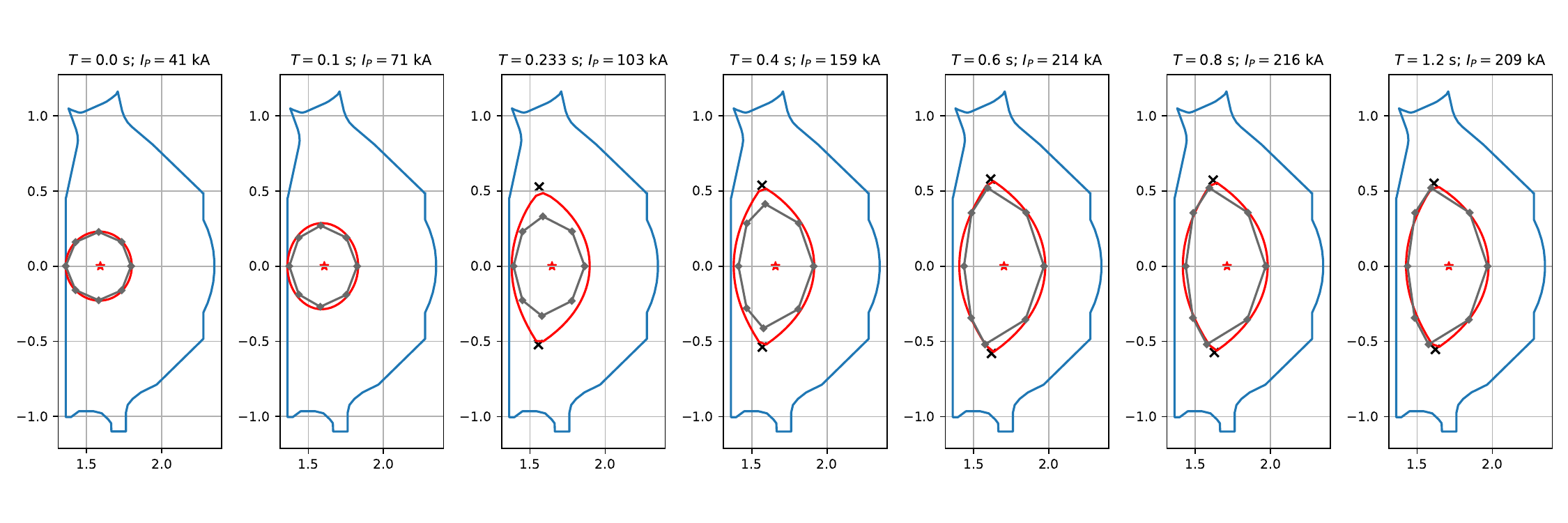}
\includegraphics[width=\textwidth]{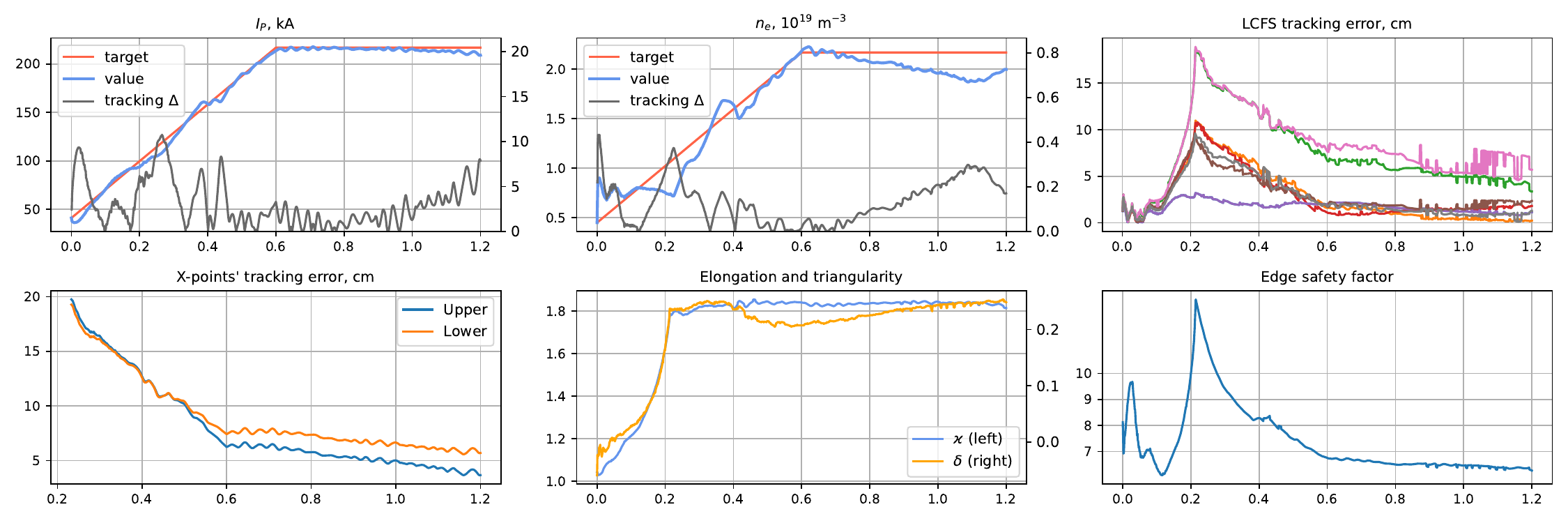}
\caption{\label{fig:lowip}Test episode with a low-current target: $I_p^{\rm t}=217$~kA,
$\bar n_e^{\rm t}=2.2\times10^{19}$~m$^{-3}$, $a=0.27$~m, $\varkappa=1.95$, targets reached
at 0.6~s. The plasma is diverted at 0.21~s and held to the end of the episode.}
\end{figure}

\begin{figure}
\centering
\includegraphics[width=\textwidth]{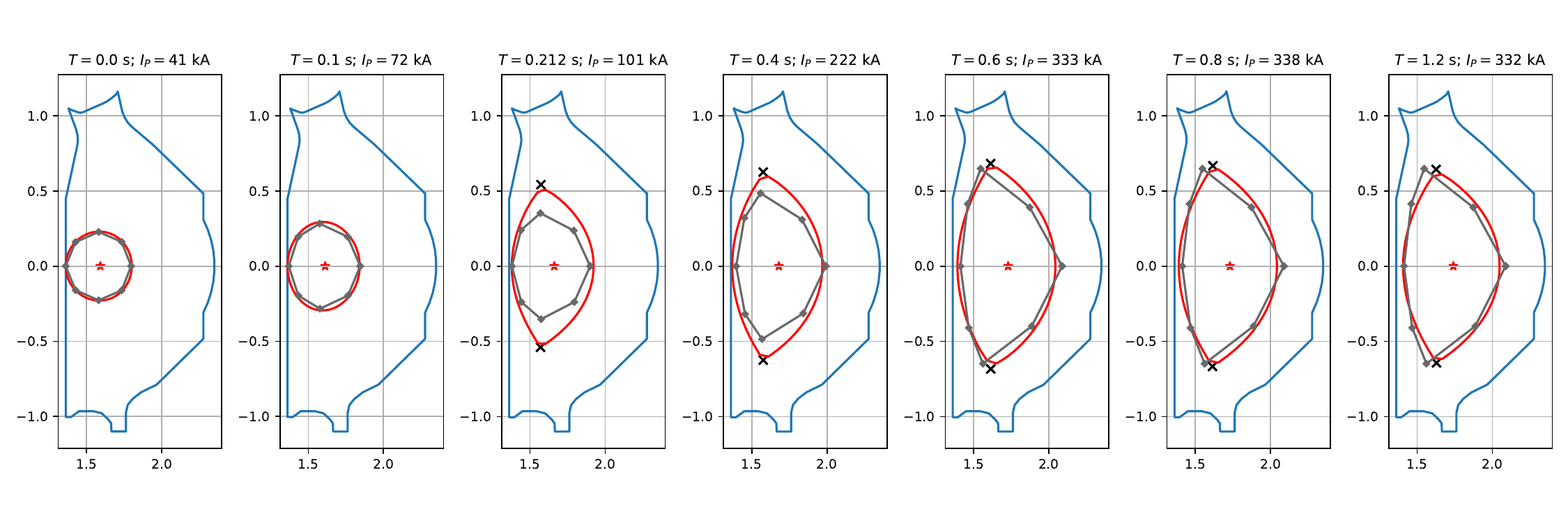}
\includegraphics[width=\textwidth]{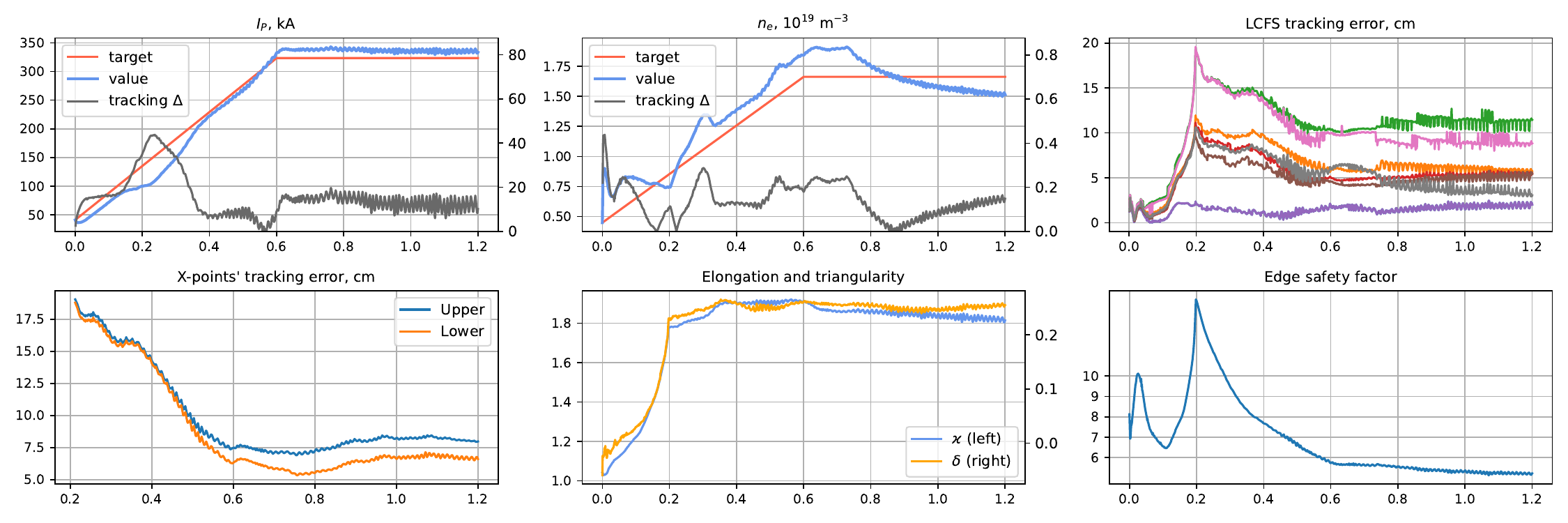}
\caption{\label{fig:highip}Test episode with a high-current, low-density target:
$I_p^{\rm t}=323$~kA, $\bar n_e^{\rm t}=1.7\times10^{19}$~m$^{-3}$, $a=0.34$~m,
$\varkappa=1.92$. The plasma is diverted at 0.20~s and held to the end of the episode;
the current settles 13~kA above its target.}
\end{figure}

\begin{figure}
\centering
\includegraphics[width=\textwidth]{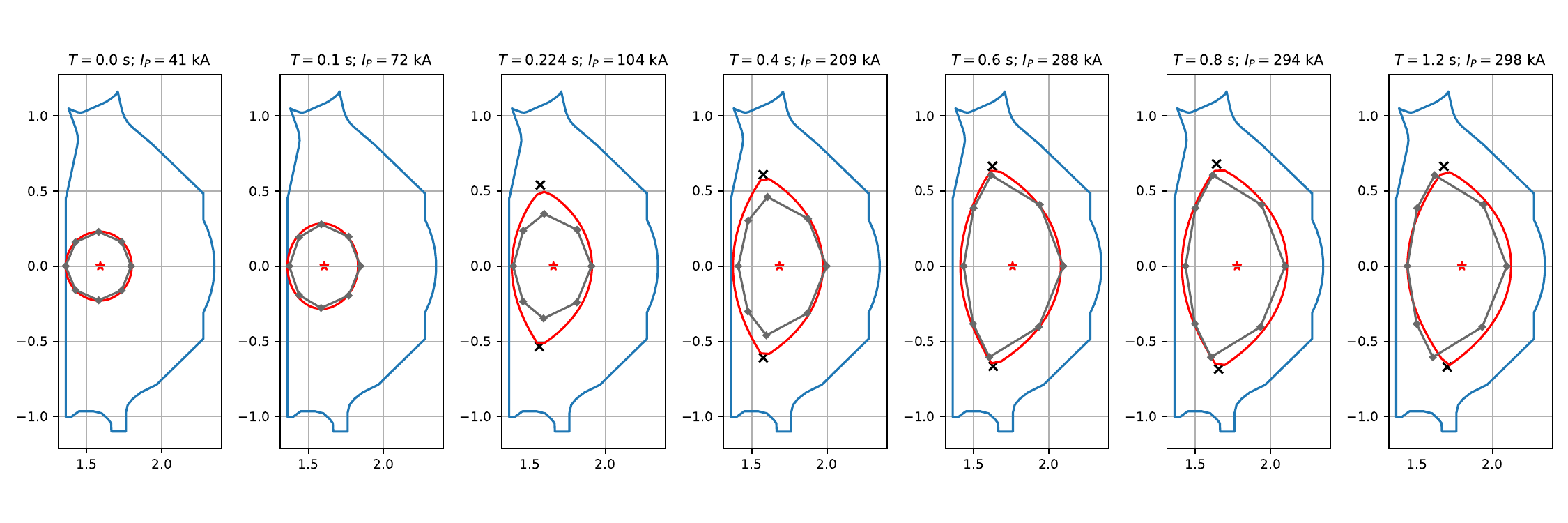}
\includegraphics[width=\textwidth]{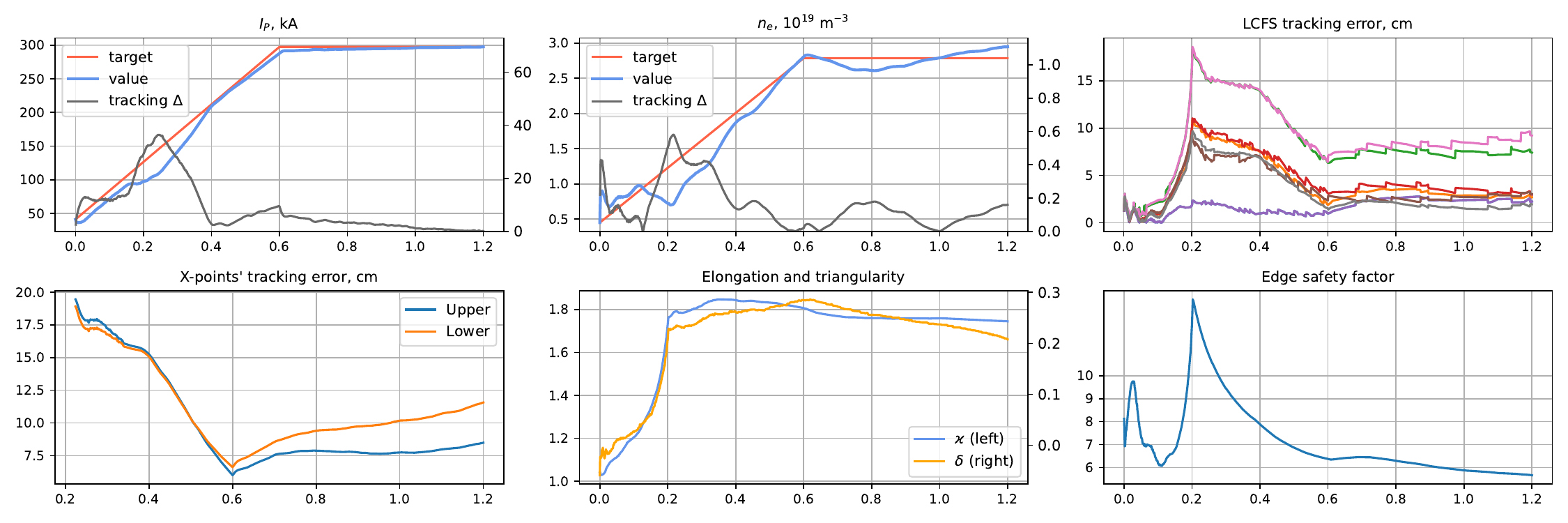}
\caption{\label{fig:highipne}Test episode with a high-current, high-density target, the
target relevant to the long-pulse targets: $I_p^{\rm t}=297$~kA,
$\bar n_e^{\rm t}=2.8\times10^{19}$~m$^{-3}$, $a=0.33$~m, $\varkappa=1.83$. The plasma is
diverted at 0.20~s and held to the end of the episode at a Greenwald fraction of 0.36.}
\end{figure}

Figures~\ref{fig:lowip}, \ref{fig:highip} and \ref{fig:highipne} show three test episodes: a low-current discharge
($I_p^{\rm t}=217$~kA at $\bar n_e^{\rm t}=2.2\times10^{19}$~m$^{-3}$), a high-current
discharge at low density ($323$~kA, $1.7\times10^{19}$~m$^{-3}$) and a high-current,
high-density discharge relevant for the long-pulse discharges ($300$~kA, 
$2.8\times10^{19}$~m$^{-3}$). On the last combination the model demonstrated
the lowest shape control accuracy. In all three episodes we use ramp up 
duration of 0.6~s, the plasma is diverted from the transition to the 
end of the 1.2~s episode without a single step back to the limiter.

Up to the transition the three episodes evolve alike. The current is ramped from
39~kA at hand-over at 0.3--0.5~MA/s, fastest for the highest target, while the
cross-section grows from the circular 0.20~m start plasma leaning on the inner wall.
Both x-points are formed at 0.20--0.21~s in all three cases, at $I_p \approx 100$~kA,
independent of the target, with the elongation already at 1.8 at
that moment.

During the flat-top ($t \ge 0.6$~s) the low-current and the high-current, high-density
episodes hold the current within 2--3~kA of the target (a steady bias of $-2$~kA), the
density within $0.1$--$0.15\times10^{19}$~m$^{-3}$, the geometric centre within 0.5--2~cm,
the eight boundary points within 2.4--3.9~cm on average (7--10~cm at the worst point) and
the x-points within 5--10~cm.

The operational limits are respected with margin in all three episodes. The largest coil
current is 10.4--11.1~kA (PF5) against the 12.5~kA rating, the in-vessel coil stays below
0.7~kA (2.0~kA in the high-current, low-density case), and the solenoid current spread
grows by 3.5--4.6~kA.  Finally, in all three episodes the safety factor
on axis settles at 0.5--0.6; the edge safety factor stays above 5, so the $q_{\rm edge}<2$
termination never approaches (see section~\ref{sec:discussion}).

\clearpage
\subsection{Wall disconnect optimization}
\begin{figure}
\centering
\includegraphics[width=\textwidth]{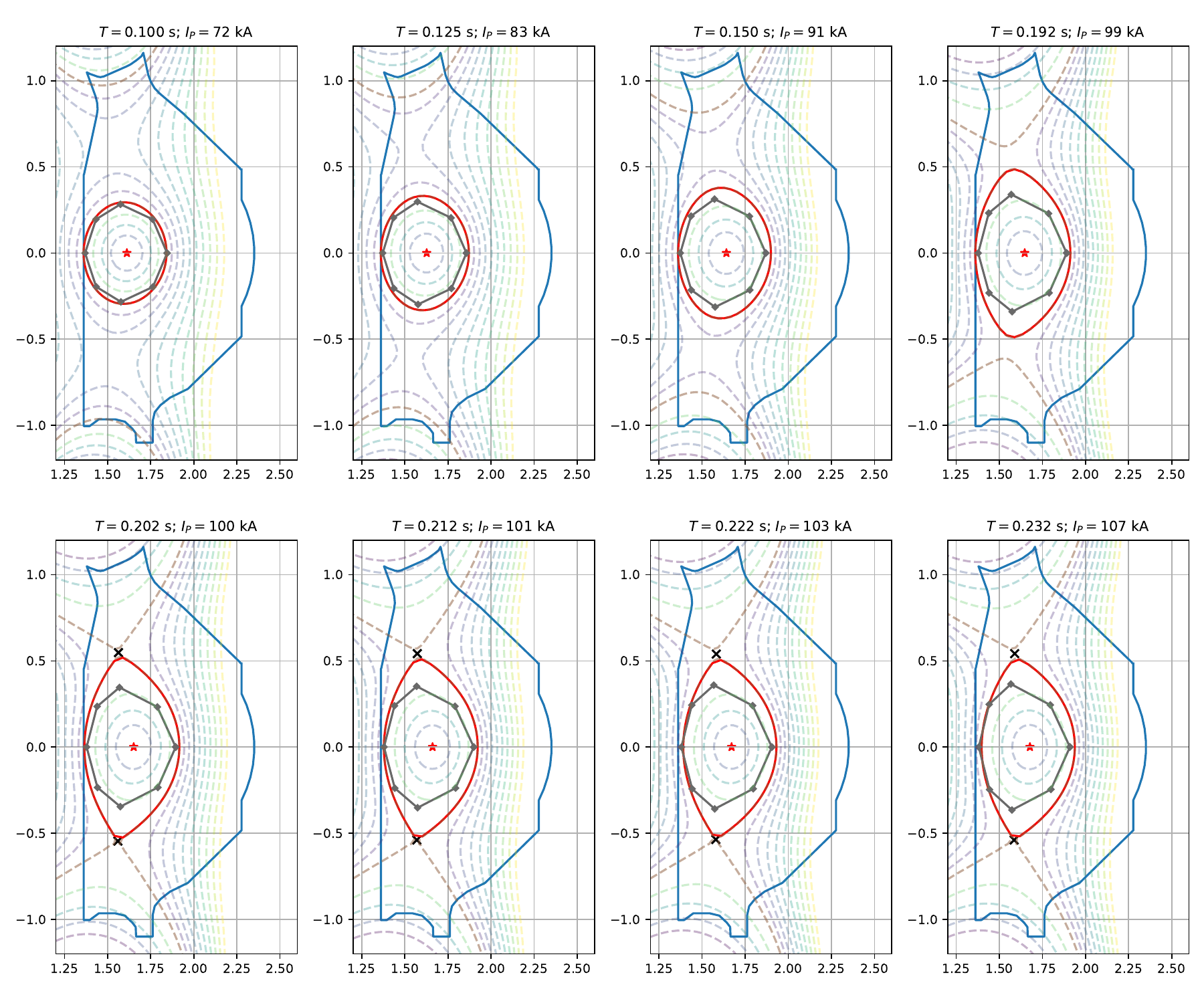}
\caption{\label{fig:transition}The evolution of plasma around limited
to diverted mode transition}
\end{figure}

As discussed earlier, the transition to the diverted configuration is
highly rewarded. This makes the other reward components more loose around
the transition (except those accompanied with termination criteria).
In the observed behavior the model elongates the plasma much faster
than piecewise linear ramp-up stipulates (Figure~\ref{fig:transition})
The model demonstrates the similar plasma evolution around the transition
with much higher elongation as the target requests and larger plasma volume. This
behavior shows up in almost all the model checkpoints with various
hyperparameters and random seeds. We argue, that the model optimizes
the smoothness of the transition and the plasma stability. 
This may require a special shape.

\clearpage
\subsection{Estimating the area of expertise}
\label{sec:eval}

To test the controller and determine its capabilities we have run an
extensive test with 1024 different targets (shape, $I_P$ and $n_e$). We sampled
space-filling set of scenarios from a low-discrepancy scrambled Sobol sequence in the 11-dimensional
parameters space (the four squarenesses, $I_p$, $\bar n_e$, $R_{\rm in}$, $R_{x1}$,
$R_{x2}$, $a$ and $\varkappa$), with every target reached at 0.6~s.
Each scenario is run once with the deterministic policy, and we recorded the episode length,
the time spent in the diverted phase, the termination reason and the full state history. A scenario
is considered successful (binary metric) if it reached the diverted mode.
We've got 876 (85.5\%) success episodes: Table~\ref{tab:sobol}. 

\begin{figure}
\centering
\includegraphics[width=0.6\textwidth]{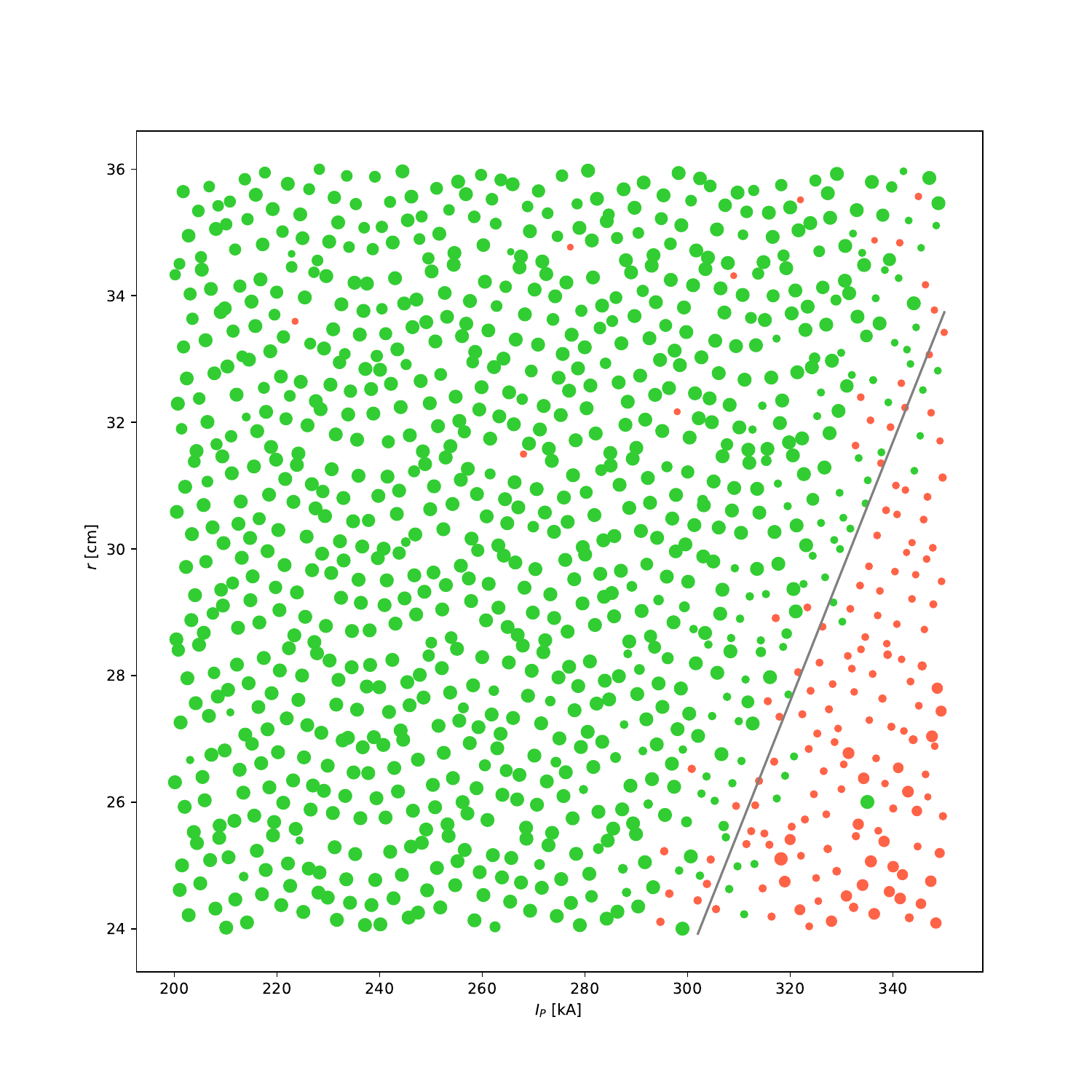}
\caption{\label{fig:sobol}Test on 1024 Sobol scenarios. Color = plasma
successfully changed mode to diverted, size = episode length (the majority of
simulations lasted full 1200 steps)}
\end{figure}

The combination of target plasma current and minor radius is by far the dominant
predictor of success (Figure~\ref{fig:sobol}). The discharge is likely successful if $r > 0.20448\cdot I_P - 37.83$
($I_P$ is in kA, $r$ is in cm).

\begin{table}
\caption{\label{tab:sobol}Test on 1024 Sobol scenarios, by target plasma current.}
\centering
\begin{tabular}{@{}lccccc}
\toprule
 & $n$ & divertor formed & held $\ge$ 0.1 s & held $\ge$ 0.5 s & held to 1.2 s\\
\midrule
$I_p^{\rm t}$ 200--250 kA & 341 & 1.00 & 0.98 & 0.85 & 0.73\\
$I_p^{\rm t}$ 250--300 kA & 341 & 0.98 & 0.95 & 0.84 & 0.74\\
$I_p^{\rm t}$ 300--350 kA & 342 & 0.59 & 0.38 & 0.33 & 0.28\\
all & 1024 & 0.86 & 0.77 & 0.67 & 0.58\\
\bottomrule
\end{tabular}
\end{table}

\clearpage
\section{Discussion, Possible Next Steps and Applications}
\label{sec:discussion}

We have shown that a single goal-conditioned RL controller can execute an EAST ramp-up
scenario in a free-boundary equilibrium plus transport simulator from a small limited
plasma to a held double-null configuration, tracking sampled current, shape
targets and electron density by a single controller, and that it does so for
a large majority of the scenarios in an 11-dimensional
target space. The controller is a complete closed-loop scenario executor: given the
target waveforms it produces coil-current and fueling commands at 1~kHz with no
scenario-specific tuning, and its test over the target space gives the scenario designer
a map of what is reachable---here, that everything below 300~kA in a 0.6~s ramp is
reliably reachable.

The limitations include the control  quality. It is adequate for the scenario
design, but not directly applicable to the real time control. However the
control quality can be improved with longer training and more aggressive margins.
For example, x-point accuracy is about 8~cm and the current accuracy is $\approx10$~kA
during the flat-top in some regimes. A conventional controller in a
Plasma Control System can achieve a better performance.

An optimization procedure run on top of this controller (ML or not) may produce
a higher level controller which can reach more high-level plasma performance
metrics. Ultimately, the simulator can be coupled with heating controller
and target the energy gain.

\appendix

\section{Observations}
\label{app:obs}

\begin{table}[H]
\caption{\label{tab:obs}Inputs of the policy network (109 values). ``offset'' means the
quantity is given as achieved minus target; ``derivative'' means its one-step finite
difference is appended. The critic receives the same inputs plus actions.}
\centering
\begin{tabular}{@{}lll}
\toprule
Input & Size & Form\\
\midrule
Goal: target centre, 8 boundary points, $I_p$, $\bar n_e$ & 20 & raw\\
Plasma current & 1 & raw, derivative\\
Coil currents (13 coils) & 13 & raw, derivative\\
Boundary points (8, $R$ and $Z$) & 16 & offset, derivative\\
Limiter contact point & 2 & raw\\
Lower and upper x-point & 4 & offset\\
Diverted flag & 1 & raw\\
Flux imbalance between nulls $\psi_{x1}-\psi_{x2}$ & 1 & raw\\
Separatrix flux & 1 & raw\\
Normalised flux at the 8 target boundary points & 8 & raw\\
$\ell_i$, $q_{\rm axis}$, $q_{\rm edge}$, $\beta_N$ & 4 & raw\\
Loop voltage, plasma resistance & 2 & raw\\
Average electron and ion temperature & 2 & raw\\
Line-averaged density & 1 & offset\\
Fueling level & 1 & raw\\
Instability growth increment, VDE estimate & 2 & raw\\
\bottomrule
\end{tabular}
\end{table}

\section{Reward components}
\label{app:reward}

\begin{table}[H]
\caption{\label{tab:reward}Reward components of the EAST ramp-up task with the margin
$\gamma$ of their Lorentzian transform (the error at which the component is worth one
half). All components have unit weight in the smooth minimum of
equation~(\ref{eq:softmin}). Phase membership: L = limited, D = diverted.}
\centering
\begin{tabular}{@{}lp{1.6in}ll}
\toprule
Component & Error $d$ & $\gamma$ & Phase\\
\midrule
plasma current & $|I_p - I_p^{\rm t}|$ & 50 kA & L, D\\
density & $|\bar n_e - \bar n_e^{\rm t}|$ & $2.0\times10^{19}$ m$^{-3}$ & L, D\\
centre & $\|C - C^{\rm t}\|$ & 0.15 m & L, D\\
boundary shape & softmax$_i\|p_i - p_i^{\rm t}\|$ (8 points, relative to the centre) & 0.6 m & L, D\\
flux at target points & softmax$_i|\psi_N(p_i^{\rm t}) - 1|$ & 0.5 & L, D\\
limiter contact height & $|Z_{\rm lim} - Z^{\rm t}|$ & 0.3 m & L\\
lower / upper x-point & $\|x_{1,2} - x_{1,2}^{\rm t}\|$  & 0.3 m & D\\
double-null balance & $|\psi_{x1} - \psi_{x2}|$ & 0.3 & D\\
CS current spread & $\Delta I_{\rm CS}(t) - \Delta I_{\rm CS}(0)$ & 9 kA & L, D\\
IC current & $|I_{\rm IC}|$ & 5 kA & L, D\\
PF current limits & $\max_c |I_c|/I_c^{\max}$, piecewise: up to 95\% $\rightarrow 1$;
100\% $\rightarrow 0.05$;\\ & 95\% to 100\% linear decrease from 1 to 0.05 & -- & L, D\\
\bottomrule
\end{tabular}
\end{table}

\section{Termination conditions}
\label{app:term}

\begin{table}[H]
\caption{\label{tab:term}Termination conditions by phase. The episode ends on the first
condition that holds; its name is recorded as the termination reason.}
\centering
\begin{tabular}{@{}lll}
\toprule
Condition & Threshold & Phase\\
\midrule
current tracking & $|I_p - I_p^{\rm t}| > 50$ kA & L, D\\
current collapse & $I_p < 20$ kA & L\\
minor radius & $a < 0.2$ m & L, D\\
edge safety factor & $q_{\rm edge} < 2$ & L, D\\
CS current spread & growth $> 9.5$ kA & L, D\\
boundary shape & $\max_i \|p_i - p_i^{\rm t}\| > 0.2$ m & L, D\\
wall contact outside the inner wall & $R_{\rm lim} > 1.5$ m & L\\
loss of divertor & $-$ & D\\
x-point tracking & $\|x_{1,2} - x_{1,2}^{\rm t}\| > 0.2$ m & D\\
double-null imbalance & $|\psi_{x1} - \psi_{x2}| > 0.1$ & D\\
time limit (truncation) & $t \ge 1.2$ s & L, D\\
\bottomrule
\end{tabular}
\end{table}

\section{Hyperparameters}
\label{app:hyper}

\begin{table}[H]
\caption{\label{tab:hyper}Agent and training hyperparameters.}
\centering
\begin{tabular}{@{}p{2.5in}p{3in}}
\toprule
Algorithm & SAC objective, TQC critic, automatic entropy tuning\\
Critics / quantiles per critic / atoms dropped & 3 / 25 / 3 of the pooled 75\\
Networks & MLP $256\times256$, SiLU\\
Input normalisation & running Welford statistics\\
Actor input / output & 109 / 10 (9 coil-current changes, 1 fueling change)\\
Critic input & 109 observations + 10 actions\\
Auxiliary head & 6 shape scalars, mean-squared loss, weight 0.1\\
Discount $\gamma$ / Polyak $\tau$ & 0.99 / 0.005\\
Learning rates (actor, critic, temperature) & $3\times10^{-5}$\\
Batch / replay size & 1024 / $10^6$\\
Gradient updates per simulator step & 1\\
Training budget & $4\times10^6$ updates\\
Simulator instances & 50 \\
Control step / episode & 1 ms / 1.2 s (1200 steps)\\
Fueling level range / change per step & $[5\times10^{3}, 2\times10^{4}]$ / $\pm 250$ \\
\bottomrule
\end{tabular}
\end{table}

\end{document}